\documentclass[preprintnumbers,amsmath,11pt,amssymb,floatfix,superscriptaddress,nofootinbib]{article}

\def\mytitle#1{\setcounter{equation}{0}
\setcounter{footnote}{0}
\begin{flushleft}\Large\textbf{#1}\end{flushleft}
\vspace{0.20cm}}
\def\myname#1{\leftline{{\large #1}}\vspace{-0.13cm}}
\def\myplace#1#2{\small\begin{flushleft}\textit{#1}\\
\texttt{#2}\end{flushleft}}

\usepackage{graphicx}
\usepackage{amsmath}
\begin{document}
\mytitle{Scale Factor - Cosmic Time Relation and Occurrences of Future Cosmic Singularities with Different Redshift Parametrizations of Dark Energy EoS-s}
\myname{$Swetalina~~Bhowmik^{*}\footnote{swetalina099@gmail.com}$ and $Ritabrata~~ Biswas^{\dagger}\footnote{biswas.ritabrata@gmail.com}$}
\myplace{Department of Mathematics, The University of Burdwan, Golapbag Academic Complex, City : Burdwan-$713104$, Dist. : Purba Barddhaman, State : West Bengal, India.}{}
\begin{abstract}
While modelling our late time cosmically accelerated universe, it is popular to involve different dark energy models, the equation of state of which can be taken as a function of the redshift and some unknown parameters. Barboza and Alcaniz have proposed one of a such kind of dark energy's EoS model. We use some new parametrizations like Feng, Shen, Li, Li I and II and Polynomial parametrizations to get more accurate concepts about the fate of our expanding universe. We try to find how the hypothesis of the fate of our universe behaves in the above background of dark energy models. Possibilities of occurrences of future cosmic singularities are studied. 
\end{abstract}
Our cosmological model is built upon the assumptions that it is homogeneous and isotropic on large scale. The metric for such a cosmic model can be geometrically represented by the Friedmann-Lemaitre-Robertson-Walker (FLRW) metric given by 
\begin{equation}
ds^2 = -c^2 dt^2 + a^2(t) \left[ \frac{dr^2}{1-k_0r^2}+r^2(d\theta^2+sin^2\theta d\phi^2)\right] ,
\end{equation}
where $k_0)$ $(-1$, $0$, $1)$ specifies the open ($=-1$), flat ($=0$) and closed ($=1$) universes respectively and the dynamic nature of our universe is supposed to be governed by the Einstein's field equations $G_{\mu\nu}=\frac{8\pi G}{c^4} T_{\mu\nu}$ where $G_{\mu\nu}$ depicts the information of the geometric part of the space-time and $T_{\mu\nu}$ is the energy-momentum part signifying the properties of the matter distribution in the concerned space-time. This energy-momentum tensor is constituted of contributions from a large number of different matter fields. Even if one is able to know the precise structure of the contributions of every of such fields and the equations of motion governing the corresponding field, the correct description of the energy-momentum tensor becomes complicated. Predictions of the  occurrences of different past and future singularities in the universe from the Einstein's equations can be done. Rather than exactly pointing out the singularities, it is quiet easier to obtain some physically reasonable inequalities for the energy-momentum tensor. Two of such noticeable singularities are weak energy condition \footnote{The energy-momentum tensor at each $p\in M$ obeys the inequality $T_{ab}W^{-a}W^{-b}\geq 0$ for any time like vector $W\in T_p$ with $G=C=1$, for matter distribution given by Trace[$-\rho$,$p$,$p$,$p$], the condition simply becomes $p+\rho\geq0$. The critical case $p+\rho=0$ is called the phantom barrier.} and strong energy condition \footnote{$3p+\rho\geq0$, the critical case $\rho+3p=0$ is popularly known to be the quintessence barrier.}. Since $1995$, two different collaborative teams of distant supernova search have observed that the distant SNeIa supernova are more redshifted. These data predict the fact of late time cosmic acceleration. To justify such a repulsive negative pressure responsible behind this phenomenon, it was hypothesised by a huge part of cosmologists/astrophysicists that a homogeneous fluid / energy is permeated all over in the universe which is responsible for such an accelerated expansion. The name of this fluid was popularly coined as dark energy/quintessence. Among many probable models of dark energy(DE hereafter), the redshift parametrization methods of the equation of state (EoS hereafter) parameter are popular.

Two conventional families \cite{ref18} of redshift parametrizations of EoS are, \\
(i) Family I : $\omega(z) = \omega_0 + \omega_1 \frac{1}{(1+z)^n} $ and \\
(ii) Family II: $\omega(z) = \omega_0 + \omega_1 \frac{1}{1+z^n} $ \\
where, $z$ is redshift, $\omega_0$ and $\omega_1$ are two undecided parameters, $n$ is a natural number. Some particular ‘$n$’-cases for both the families I and II are very popularly studied in literature and are known as :\\
\large{\textbf{(i)Linear Parametrization}, (for $n=0$ in family II) EoS is 
$\omega(z)=\omega_0+\omega_1(z)$ \cite{ref4}.\\
\large{\textbf{(ii)CPL Parametrization}},(After Chevallier, Polarski and Linder; $n=1$ for families I and II) EoS is $\omega(z)=\omega_0+\omega_1(\frac{z}{1+z})$ \cite{ref1}. \\
\large{\textbf{(iii)JBP Parametrization}}, (After Jassal, Bagala and Padmanabhan; for $n=2$ in family II) EoS is $\omega(z)=\omega_0+\frac{\omega_1 z}{(1+z)^2}$ \cite{ref6}.\\
\large{\textbf{(iv)Log or Efstathiou Parametrization}}, EoS is $\omega(z) = \omega_0 +\omega_1 ln(1+z)$ \cite{ref7}; which is valid for $z<4$.\\
\large{\textbf{(v)ASSS Parametrization}}, (After  Alam, Sahni, Saini and Starobinski \cite{ref8, ref9}) EoS is \\$\omega(z)=\left\lbrace -1+\frac{(1+z)}{3}\frac{A_1+2A_2(1+z)}{A_0+2A_1(1+z)+A_2(1+z)^2}\right\rbrace $. \\
\large{\textbf{(vi)Upadhye Ishak Steinhardt Parametrization}}, EoS is $\omega(z)= \omega_0+\omega_1z~~~~~if z<1$ and                                                         $\omega(z)= \omega_0+\omega_1~~~~~~~~if z\geq 1$ \cite{ref5}. \\
\large{\textbf{(vii)Hannestad M{\"o}rtsell Parametrization}}, EoS is $\omega(z)=\omega_0\omega_1\large{\frac{a^p+a^{p}_{s}}{\omega_1 a^p+\omega_0 a^{p}_{s}}}=\frac{1+\left(\frac{1+z}{1+z_s} \right)^p }{\omega_0^{-1}+\omega_{1}^{-1}\left(\frac{1+z}{1+z_s} \right)^p}$ \cite{ref13}. \\
\large{\textbf{(viii)Lee Parametrization}}, EoS is $\omega(z)$ as $\omega(z)=\omega_r\frac{\omega_0~exp(px)~+~exp(px_c)}{exp(px)~+~exp(px_c)}$ \cite{ref14} .\\
\large{\textbf{(ix)Barboza Alcaniz Parametrization:}}
\\The BA \cite{ref19} EoS is 
\begin{equation}
\omega(z)=\omega_0+\omega_{1}\frac{z(1+z)}{1+z^{2}},
\end{equation}
$\omega_{0}$ is the EoS at present time $z=0$ and $\omega_{1} = \frac{d\omega}{dz}$ at $z=0$. These information give a measurement of time dependence of this DE EoS.
For this parametrization, the bounds in $\omega_{0}-\omega_{1}$ plane are given as:-\\
\textit{For quintenssence:} \\ $-1\leq\omega_{0}-0.21\omega_{1}$  and   $\omega_{0}+1.21\omega_{1}\leq 1$;    in case of $\omega_{1}>0$ \\and   $-1\leq\omega_{0}+1.21\omega_{1}$ and $\omega_{0}-0.2\omega_{1}\leq 1$; in case of $\omega_{1}<0$ \cite{ref19}
\\ \textit{For phantom:} \\$\omega_1<-\frac{(1+\omega_0)}{1.21}$   (when $\omega_1>0$) \\and $\omega_1>\frac{(1+\omega_0)}{0.21}$   (when $\omega_1<0$) \cite{ref19} \\
\large{\textbf{(x)Feng Shen Li Li Parametrization:}}
\\To surpass the divergence of the CPL model (for $z\rightarrow-1$) Feng, Shen, Li and Li \cite{ref15} suggested following interesting relations: 
\begin{equation}
\left.  \begin{aligned}
\\FSLL~~I:   \omega(z)=\omega_0+\omega_1\frac{z}{1+z^2} 
\\FSLL~~II:   \omega(z)=\omega_0+\omega_1\frac{z^2}{1+z^2}
\end{aligned}\right\rbrace
\end{equation}

 Here, $\omega_0=\omega(0)$ and $\omega_1=\frac{d\omega}{dz}|_{z=0}$. In the 1st case, $\omega(\infty)=\omega_0$ and it reduces to $\omega(z)\approx\omega_0+\omega_1 z$ at low $z$. Again, for the second one, $\omega(\infty)=\omega_0+\omega_1$ and it yields $\omega(z)\approx\omega_0+\omega_1 z^2$ at low redshifts.\\
\large{\textbf{(xi)Polynomial Parametrization:}}
\\Sendra and Lazkoz once proposed polynomial parametrization in an expansion in powers of $(1+z)$, which is given as follows \cite{ref16,ref17}
\begin{equation}
\omega(z)=-1+c_1\left(\frac{1+2z}{1+z} \right)+c_2\left(\frac{1+2z}{1+z} \right)
\end{equation}. Here, $c_1=(16\omega_0-9\omega_{0.5}+7)/4$ and $c_2=-3\omega_0+(9\omega_{0.5}-3)/4$; the values of the EoS are $\omega_0$ and $\omega_{0.5}$ at $z=0$ and $z=0.5$ respectively.

The \textbf{CPL} and \textbf{linear} models diverge for large $z$ ($z\gg0$). Here we take some new parametrizations to get a proper concept about the fate of our universe and to speculate about future cosmic singularities.

The Einstein's field equations for the metric given by the equation ($1$) are written in the form \cite{ref10}:
\begin{equation}
\left.  \begin{aligned}
\\ \frac{2\ddot{a}}{a}+\frac{\dot{a}^2+k_0}{a^2}=-\kappa p+\Lambda 
\\ 3\Big(\frac{\dot{a}^2+k_0}{a^2}\Big)=\kappa \rho+\Lambda
\end{aligned}\right\rbrace
\end{equation} with $\kappa=8\pi$.
The energy conservation equation is:
\begin{equation}
\dot{\rho}+3\frac{\dot{a}}{a}(p+\rho)=0.
\end{equation}
From (1) and field equations (5) we easily get
\begin{equation}
\ddot{a}=-\frac{\kappa}{6}(p+\rho)a+a\frac{\Lambda}{3}.
\end{equation}
This is the Raychaudhuri equation \cite{ref11,ref12} which provides the cosmic acceleration that is governed by forces on the right hand side of it. So we obtain
\begin{equation}
-\frac{k_0}{2}=\frac{1}{2}\dot{a}-\Big(\frac{4\pi}{3}\rho+\frac{\Lambda}{6}\Big)a^2
\end{equation}
Now a function $M(\rho)$ is defined with the help of the equations $(1),(5),(6)$ and $(8)$ for finding the general solution of the system as:
\begin{equation}
M(\rho) = exp \left[ \int \frac{dp}{p+\rho}\right] > 0.
\end{equation}.
We consider the pressure $p$ as a function of density $\rho$ and obtain 
\begin{equation}
\frac{dM(\rho)}{d\rho}=\frac{M(\rho)}{p+\rho} > 0.
\end{equation}
We now write the conservation equation $(6)$ as;
\begin{equation}
\frac{d}{dt}[lnM(\rho)+ln a^3]=0\Rightarrow M(\rho)a^3=m_0.
\end{equation}

In this letter, we will calculate the relations between $a(t)$ and $t$ for BA, FSLL I, FSLL II and polynomial parametrizations one by one in the next parts. We will graphically interpret these relations then. Lastly we will briefly discuss the results achived and draw a conclusion.


We use the EoS of BA parametrization in the expression relating the mass and density, i.e., in equation  $(9)$ and will get the relation of $M$, $\rho$ and $z$ as
\begin{equation}
m_0a^{-3}=M=\rho^{\frac{1}{1+\omega_0+\omega_1\frac{z(1+z)}{1+z^{2}}}}\Rightarrow \rho=(m_0a^{-3})^ {\left\lbrace {1+\omega_0+\omega_{1} \frac{z(1+z)}{1+z^{2}}} \right\rbrace }
\end{equation}
Again using equations $(5)$ and $(7)$
\begin{eqnarray}
\dot{a}^2=2\left\lbrace \frac{4\pi}{3}(m_0a^{-3})^{\left\lbrace 1+\omega_0+\omega_{1}\frac{z(1+z)}{1+z^{2}}\right\rbrace }+\frac{\Lambda}{6}\right\rbrace a^2-k_0 \nonumber \\\Rightarrow \left(\frac{da}{dt}\right)=\left[2\left\lbrace \frac{4\pi}{3}(m_0a^{-3})^{\left\lbrace 1+\omega_0+\omega_{1}\frac{z(1+z)}{1+z^{2}}\right\rbrace }+\frac{\Lambda}{6}\right\rbrace a^2-k_0 \right]^{\frac{1}{2}} 
\end{eqnarray}
Writing $t$ with respect to $a(t)$ we obtain
\begin{eqnarray}
\int _{t_0}^{t} dt=\int _{a(t_0)}^{a} \left[2\left\lbrace \frac{4\pi}{3}(m_0a^{-3})^{\left\lbrace 1+\omega_0+\omega_{1}\frac{z(1+z)}{1+z^{2}}\right\rbrace }+\frac{\Lambda}{6}\right\rbrace a^2-k_0 \right]^{-\frac{1}{2}}da \label{eq1}
\end{eqnarray}
To obtain the analytic solution of $(10)$, we put $\omega_0=-1$, $\omega_1=0$, $m_0=1$ and get:\\For $k_0=0$
\begin{equation}
t(a)= \sqrt{\frac{3}{8\pi-1}}ln a
\end{equation}
\\For $k_0=1$ 
\begin{equation}
t(a)= \sqrt{\frac{3}{8\pi-1}}ln \left( -2a\sqrt{8\pi-1}+2\sqrt{a^2(8\pi-1)-3}\right) 
\end{equation}
\\For $k_0=-1$ 
\begin{equation}
t(a)= \sqrt{\frac{3}{8\pi-1}} \left(sinh^{-1}\left[\frac{a\sqrt{8\pi-1}}{\sqrt{3}} \right] \right)
\end{equation}

Solving above equation (\ref{eq1}) numerically, we plot graphs of $t$ vs $a(t)$ for $k_0=1,0,-1$. For $k_{0}=1$, we get the graph $1(a)$. Here we discuss the case for quintenssence and phantom era for different values of $k$ and $\omega$. 
\begin{figure}[ht!]
\begin{center}
$~~~~Fig.1(a)~~~~~~~~~~~~~~~~~~~~~~~~~~~~~~~~~~~~~Fig.1(b)~~~~~~~~~~~~~~~~~~~~~~~~~~~~~~~~~~~~~~Fig.1(c)$
\includegraphics[height=5cm, width=5.6cm]{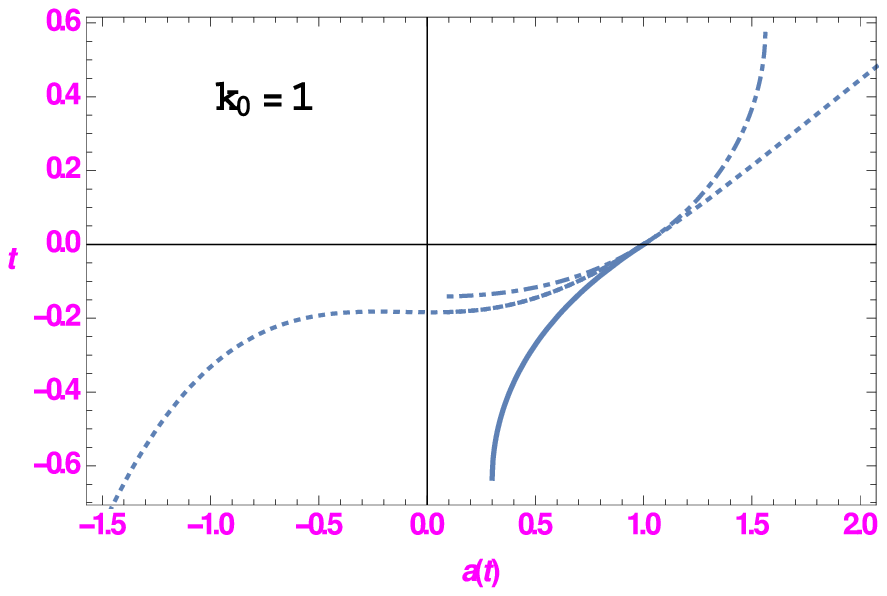}~~~~~\includegraphics[height=5cm, width=5.6cm]{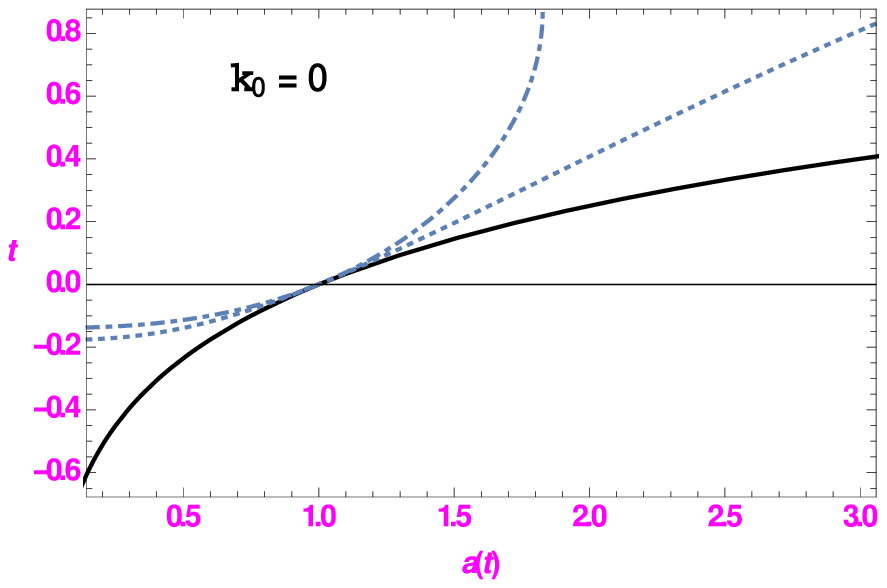}~~~~~\includegraphics[height=5cm, width=5.6cm]{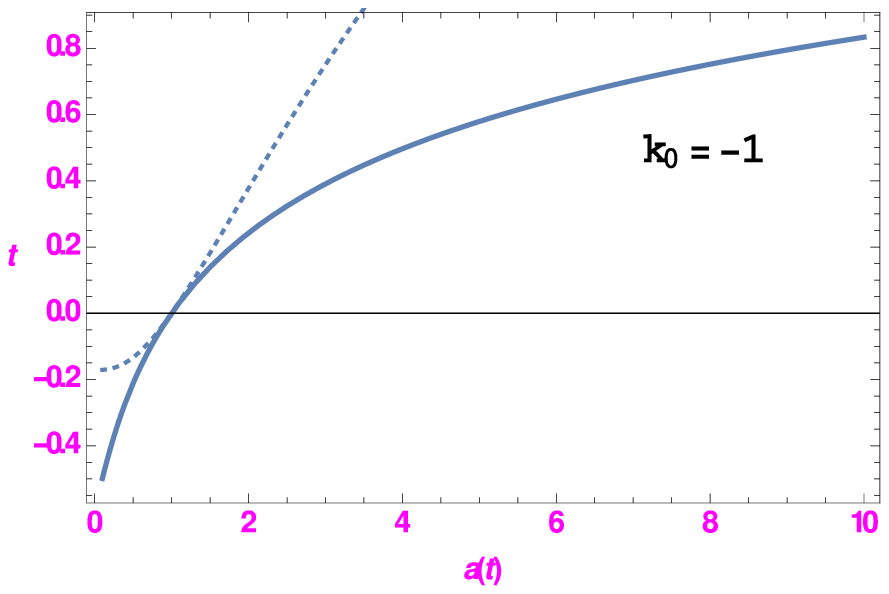}
\end{center}
Figure $1(a)$ to $1(c)$ are $t$ vs. $a(t)$ plots for BA parametrization for $k_0=1$, $k_0=0$ and $k_0=-1$ respectively. For Fig.$1(a)$: solid line stands for $\omega_0=-1$, $\omega_1=0.1$; dotted line represents $\omega_0=-0.80423$, $\omega_1=1.40845$ and the dot-dashed one stands for $\omega_0=1.006$ and $\omega_1=-0.41493775933$. For Fig.$1(b)$: solid line stands for $\omega_0=-1$, $\omega_1=0.1$; dotted line represents $\omega_0=-0.80423$, $\omega_1=1.40845$ and the dot-dashed one stands for $\omega_{0}=0.9170224481,\omega_{1}=-0.3149677893$. for Fig.$1(c)$: solid line stands for $\omega_0=-1$, $\omega_1=0.1$ and the dotted line represents $\omega_0=-0.80423$, $\omega_1=1.40845$ 
\end{figure}

The solid line states that for negative time we may have a negative (but increasing) $a(t)$, i.e., if we choose $\omega_0=-1$ and $\omega_1=0.1$, in past we may observe a deceleration. But if $\omega_0=-0.80423$ and $\omega_1=1.40845$, i.e., for dotted curve we see as $t$ increases $a(t)$ increases as well. However, when $\omega_0=1.006$ and $\omega_1=-0.41493775933$ i.e. in the dot-dashed graph we see if $t$ increases, $a(t)$ becomes asymptotic to a finite value. This third case does not allow any future cosmological singularity. However, in closed universe the $\omega_0=1.9170224481$  case does not give any physical value for $t>0$. 

We can suppose that $a(t)$ blows for increasing $t$. The same pattern is followed for the flat universe case.

For open universe, (Fig. $1(c)$) the graph where $a(t)$ becomes asymptotic to a finite value for increasing $t$ is absent. This signifies the open universe does not allow not to possess a future singularity.
 
\begin{figure}[ht!]
\begin{center}
$~~~~Fig.2(a)~~~~~~~~~~~~~~~~~~~~~~~~~~~~~~~~~~~~~Fig.2(b)~~~~~~~~~~~~~~~~~~~~~~~~~~~~~~~~~~~~~~Fig.2(c)$
\includegraphics[height=5cm, width=5.6cm]{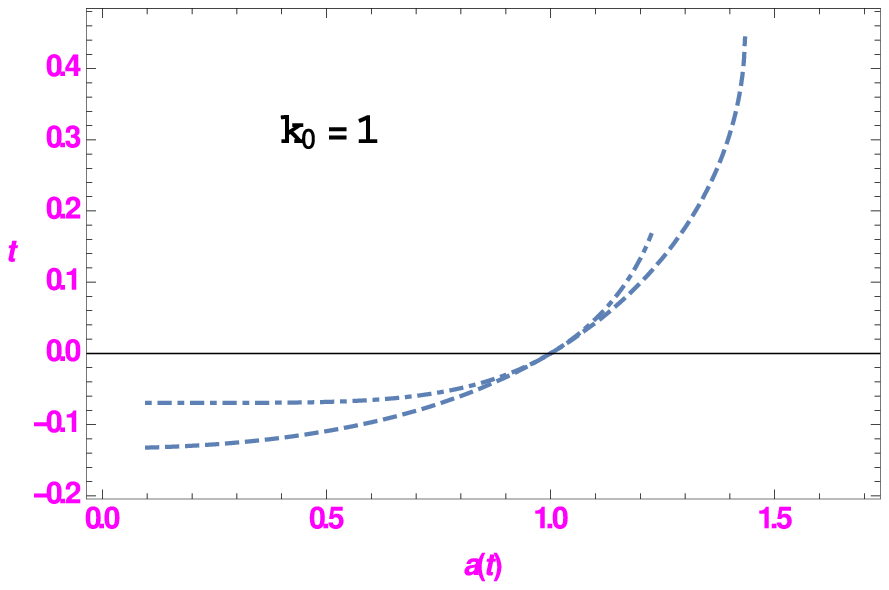}~~~~~\includegraphics[height=5cm, width=5.6cm]{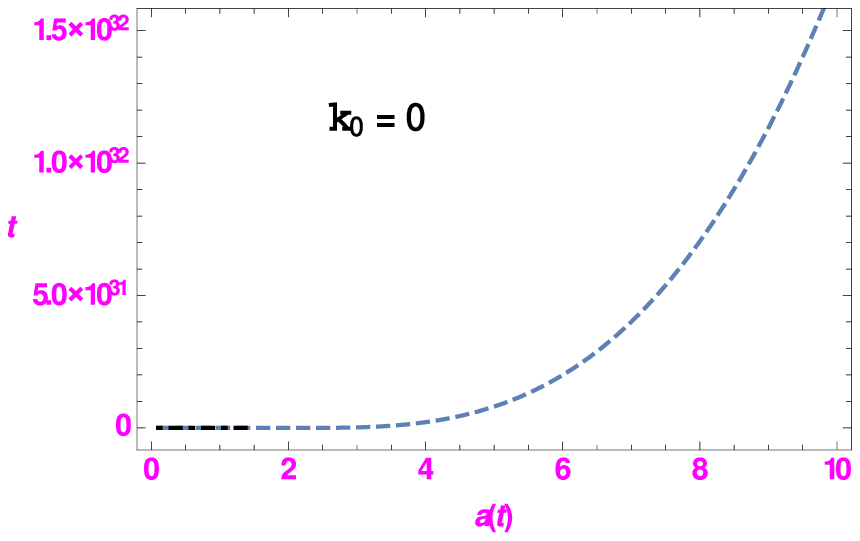}~~~~~\includegraphics[height=5cm, width=5.6cm]{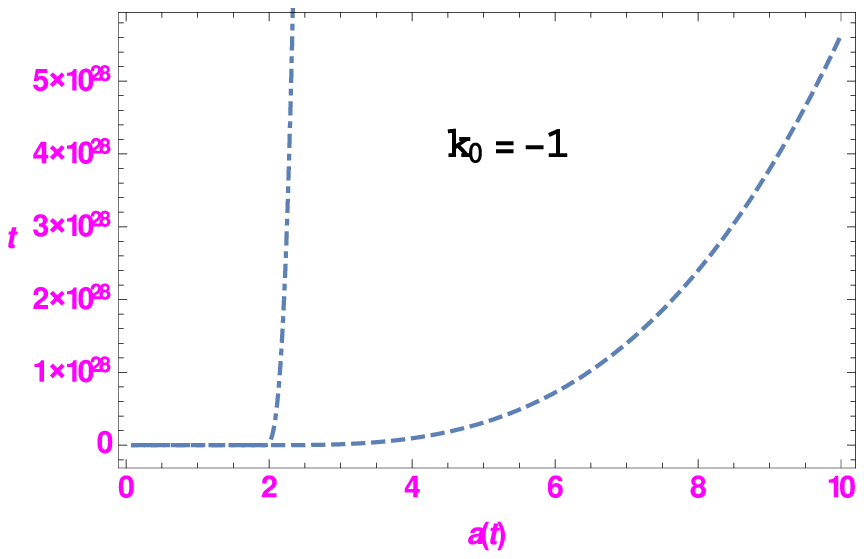}
\end{center}
Figure $2(a)$ to $2(c)$ are $t$ vs. $a(t)$ plots for BA parametrization for $k_0=1$, $k_0=0$ and $k_0=-1$ respectively. For Fig.$2(a)$: dashed line represents $\omega_0=1.8$, $\omega_1=-1.3105$ and the dot-dashed one stands for $\omega_0=1.1$ and $\omega_1=1.9$. For Fig.$2(b)$: dashed line represents $\omega_0=1.1$, $\omega_1=-1.895$ and the dot-dashed one stands for $\omega_{0}=1.1,\omega_{1}=1.4$. For Fig.$2(c)$: dashed line stands for $\omega_0=1.8$, $\omega_1=0.799$; dotted line represents $\omega_0=1.1$, $\omega_1=0.9$. 
\end{figure}

In Fig.$2(a)$ the dashed ($\omega_0=1.8$, $\omega_1=-1.3105$) and dot-dashed ($\omega_0=1.1$ and $\omega_1=1.9$) lines both are steeply increasing sensitive curves such that they become unphysical if we make a little change in the values of $\omega_0$ or $\omega_1$ even upto $5$ or $6$ decimal places. the dot-dashed line behaves quiet curious. Keeping $\omega_0$ fixed if we put $\omega_1=1.8502$ then the graph is also physical but the range of $t$ suddenly rises upto $2.5\times10^{27}$. Again, it becomes unphysical at $\omega_1=1.850235$ and remains the same upto $\omega_1=1.89899$.

In Fig.$2(b)$ the dotted line ($\omega_1=-1.895$) and the dashed line ($\omega_1=1.4$), keeping $\omega_0=1.1$ fixed, are steeply increasing. The dotted one strictly increases with the increment of $t$ after $a(t)=5.6$ and becomes unphysical if we change the value of $\omega_1$ even in $5$ decimal place. The dashed one increases for $a(t)>0.35$ and becomes unphysical if we change the value of $\omega_1$ from $1.4$ to $1.402$.

In Fig.$2(c)$ the dashed($\omega_0=1.8$, $\omega_1=0.799$) and dotted($\omega_0=1.1$ and $\omega_1=0.9$) lines are too much sensitive such that they become unphysical if a little change in the values of $\omega_0$ or $\omega_1$ has made.

Surprisingly, it is to be pointed out that if we take such values of Barboza Alcaniz parameters that phantom era is signified, we observe the values of $a(t)$ to converge to a finite value for increasing $t$ if we take closed universe. Flat and open universe cases do not however support existence of constant (or asymptotic to a finite value) $a(t)$ for infinite $t$. But for these two cases $a(t)$ is not diverging to an infinite value for increasing $t$. So, it is clear that Barboza Alcaniz does not support infinite $a(t)$ even when the parameters are signifying phantom era.


After describing the nature of expanding universe with a dark energy of type BA, we will have followed for \textbf{FSLL I} parametrization. For this case, again, the EoS of FSLL I given by equation $(9)$, we will get
\begin{equation}
m_0a^{-3}=M=\rho^{\frac{1}{1+\omega_0+\omega_1\frac{z}{1+z^{2}}}}\Rightarrow \rho=(m_0a^{-3})^{(1+\omega_0+\omega_{1}\frac{z}{1+z^{2}})}
\end{equation}
Again using equations $(5)$ and $(7)$
\begin{eqnarray}
\dot{a}^2=2\left\lbrace \frac{4\pi}{3}(m_0a^{-3})^{\left\lbrace 1+\omega_0+\omega_{1}\frac{z}{1+z^{2}}\right\rbrace }+\frac{\Lambda}{6}\right\rbrace a^2-k_0 \nonumber \\\Rightarrow \left(\frac{da}{dt}\right)=\left[2\left\lbrace \frac{4\pi}{3}(m_0a^{-3})^{\left\lbrace 1+\omega_0+\omega_{1}\frac{z}{1+z^{2}}\right\rbrace }+\frac{\Lambda}{6}\right\rbrace a^2-k_0 \right]^{\frac{1}{2}} .
\end{eqnarray}
Here, integratiating $t$ with respect to $a(t)$ we get
\begin{eqnarray}
t-t_0=\int_{a(0)}^{a} \left[2\left\lbrace \frac{4\pi}{3}(m_0a^{-3})^{\left\lbrace 1+\omega_0+\omega_{1}\frac{z}{1+z^{2}}\right\rbrace }+\frac{\Lambda}{6}\right\rbrace a^2-k_0 \right]^{-\frac{1}{2}}da \label{eq2}
\end{eqnarray}
To obtain the analytic solution of $(16)$,we put $\omega_0=-1$, $\omega_1=0$, $m_0=1$ and get:\\For $k_0=0$
\begin{equation}
t(a)= \sqrt{\frac{3}{8\pi-1}}ln a
\end{equation}
\\For $k_0=1$ 
\begin{equation}
t(a)= \sqrt{\frac{3}{8\pi-1}}ln \left( -2a\sqrt{8\pi-1}+2\sqrt{a^2(8\pi-1)-3}\right) 
\end{equation}
\\For $k_0=-1$ 
\begin{equation}
t(a)= \sqrt{\frac{3}{8\pi-1}} \left(sinh^{-1}\left[\frac{a\sqrt{8\pi-1}}{\sqrt{3}} \right] \right)
\end{equation}.

Solving the equation (\ref{eq2}) numerically we plot graphs of $t$ vs $a(t)$ for $k_0=1, 0, -1$. For $k_{0}=1$, we get the graph 3(a). Here we compare the cases for quintenssence and phantom era with BA parametrization for different values of $k$ and $\omega$. 
\begin{figure}[ht!]
\begin{center}
$~~~~Fig.3(a)~~~~~~~~~~~~~~~~~~~~~~~~~~~~~~~~~~~~~Fig.3(b)~~~~~~~~~~~~~~~~~~~~~~~~~~~~~~~~~~~~~~3(c)$
\includegraphics[height=5cm, width=5.6cm]{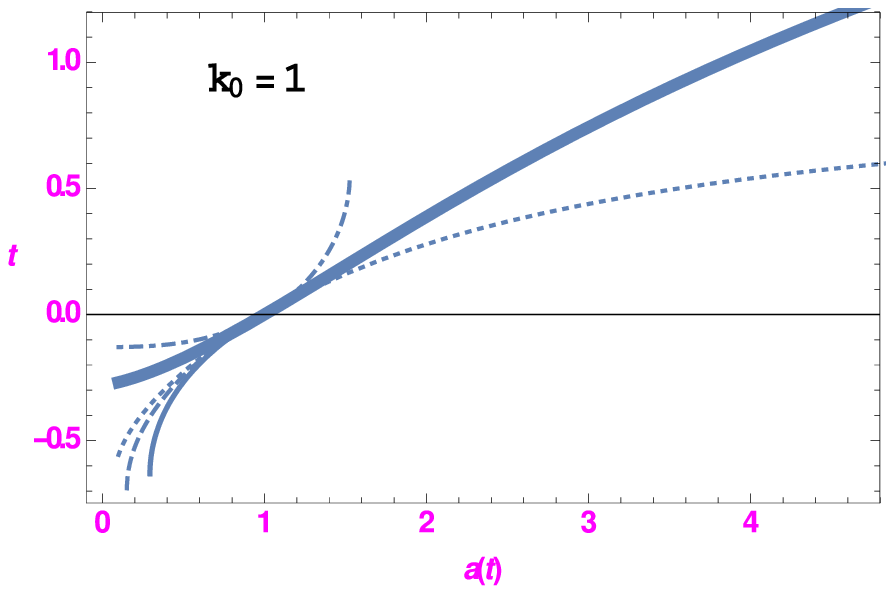}~~~~~\includegraphics[height=5cm, width=5.6cm]{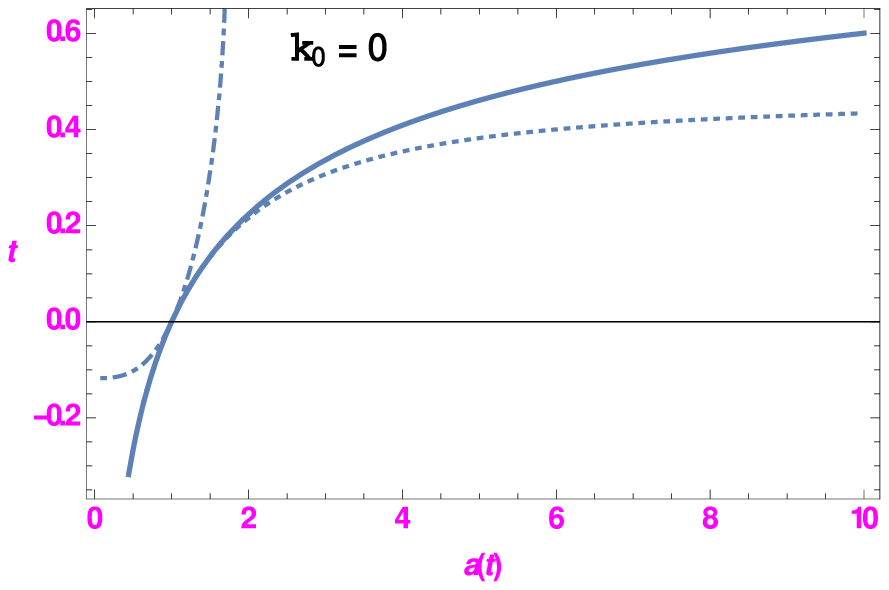}~~~~~\includegraphics[height=5cm, width=5.6cm]{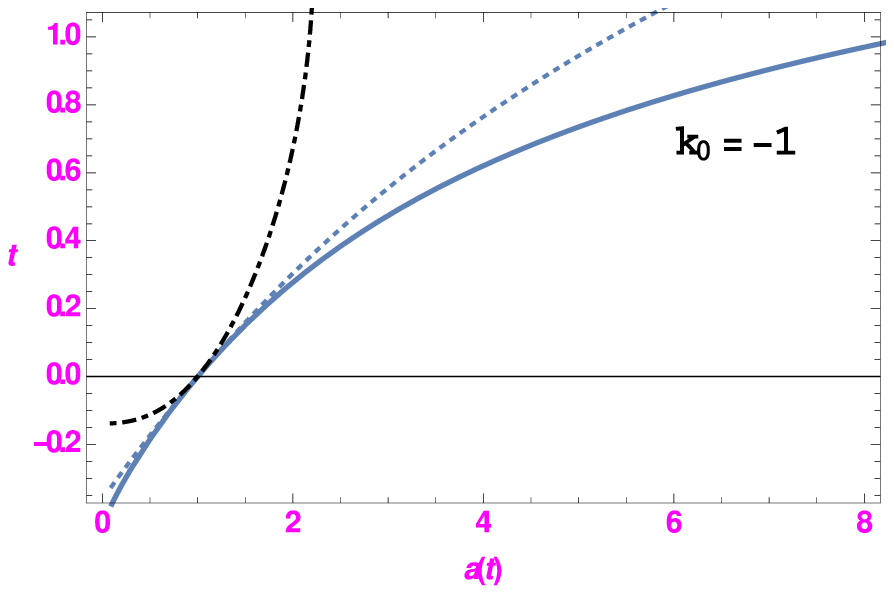}
\end{center}
Figure $3(a)$ to $3(c)$ are $t$ vs. $a(t)$ plots of FSLL I parametrization for quintessence era for $k_0=1$, $k_0=0$ and $k_0=-1$ respectively. For Fig.$3(a)$: solid line stands for $\omega_0=-1$, $\omega_1=0.2$; dotted line represents $\omega_0=-1.30423$, $\omega_1=1.40845$; dashed line represents $\omega_0=-1.40423$, $\omega_1=1.40845$; the thicker line represents $\omega_0=-0.80423$, $\omega_1=1.40845$ and the dot-dashed one stands for $\omega_0=1.006$, $\omega_1=-0.34493775933$. For Fig.$3(b)$: solid line stands for $\omega_0=-1.2$, $\omega_1=0.11$; dotted line represents $\omega_0=-0.70423$, $\omega_1=1.40845$ and the dot-dashed one stands for $\omega_{0}=1.1170224481,\omega_{1}=-0.3142677893$. For Fig.$3(c)$: solid line stands for $\omega_0=-1$, $\omega_1=0.9$; dotted line represents $\omega_0=-0.80423$, $\omega_1=0.80845$ and the dot-dashed one represents $\omega_0=0.9170224481$, $\omega_1=-0.6149677893$.
\end{figure}

In Fig. $3(a)$; the solid, dashed and dotted lines show that for negative time we get an increasing negative $a(t)$; which represents deceleration in past time. The dot-dashed one gives no cosmological singularity in future as here $a(t)$ becomes asymptotic to a finite value for increasing $t$.

In Fig. $3(b)$ and $3(c)$, if we study the increment of $a(t)$ with respect to $t$, then we note that the dotted and solid lines diverge with $a(t)$ while the dot-dashed one becomes convergent. For both the graphs the solid line become asymptotic to a finite value for increasing $t$. Here the pattern of flat and open universe is almost same.

Here in quintessence era, it has been noticed that in each cases of closed, flat and open universe ($k_0=1, 0, -1$), the dot-dashed line is always convergent with the increment of $t$ and all other curves are divergent. Surprisingly we observe that the dot-dashed lines in each graphs has been plotted for ($\omega_0>0, \omega_1<0$), which are near phantom era  while all other lines have been drawn for ($\omega_0<0, \omega_1>0$).

\begin{figure}[ht!]
\begin{center}
$~~~~Fig.4(a)~~~~~~~~~~~~~~~~~~~~~~~~~~~~~~~~~~~~~Fig.4(b)$
\includegraphics[height=5cm, width=5.6cm]{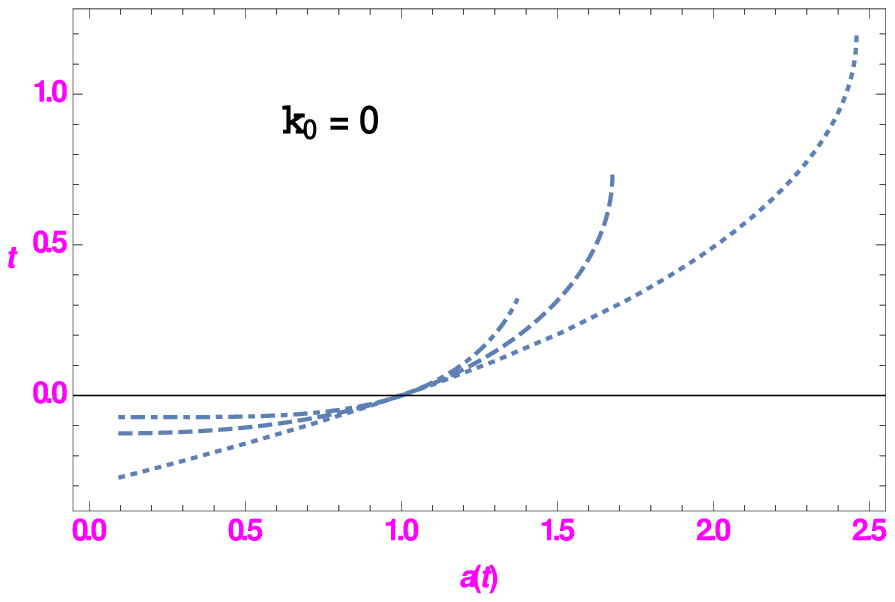}~~~~~\includegraphics[height=5cm, width=5.6cm]{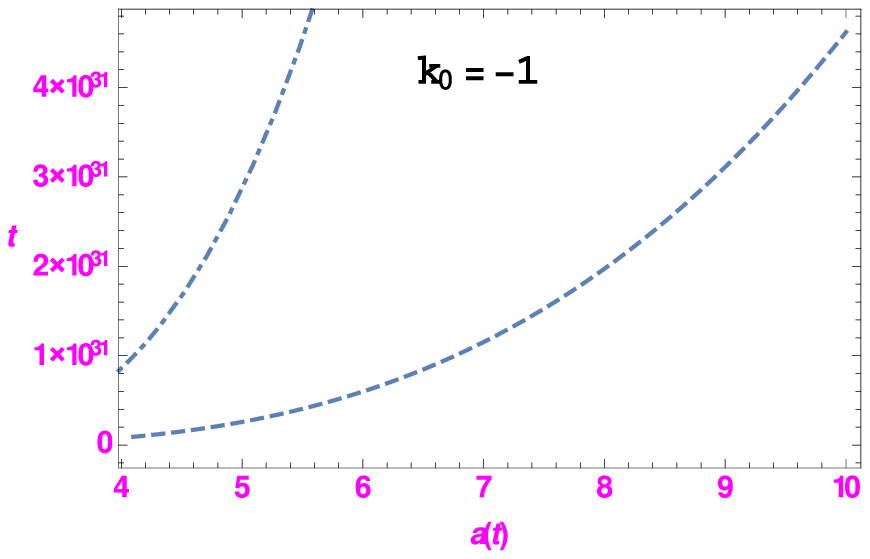}
\end{center}
Figures $4(a)$ and $4(b)$ are $t$ vs. $a(t)$ plots of FSLL I parametrization for phantom era for $k_0=1$, $k_0=0$ and $k_0=-1$ respectively. For Fig.$4(a)$: dashed line represents $\omega_0=1.7$, $\omega_1=-1.895$; dotted line stands for $\omega_0=0.7$, $\omega_1=-1.895$ and the dot-dashed one stands for $\omega_0=1.006$ and $\omega_1=-0.41493775933$. For Fig. $4(b)$: dashed line represents $\omega_0=1.8$, $\omega_1=-0.699$ and the dot-dashed one stands for $\omega_{0}=1.1,\omega_{1}=0.6$. 
\end{figure}

In fig. $4(a)$ all the dotted, dashed and dot-dashed lines converge with the increment of $a(t)$. For same $\omega_1$ ($-1.895$), we get dotted($\omega_0=0.7$) and dashed ($\omega_0=1.7$) lines. The dotted one where $a(t)$ is convergent and asymptotic to a finite value with the increment of $t$; whither both dashed and dot-dashed ($\omega_0=1.006$ and $\omega_1=-0.41493775933$) lines are absent for increasing $t$.

In fig. $4(b)$ ($k_0=-1$ i.e. for open universe) both the dashed and dot-dashed lines are neatly convergent. Convergence of the dot-dashed line is more accurate than the dashed one. This is for the data of dot-dashed one ($\omega_0=1.1, \omega_1=0.6$) which is very much alike quintessence barrier while that of the dotted one ($\omega_0=1.8, \omega_1=-0.699$) lies in phantom era.

Therefore, the FSLL I parametrization is defined in such a way that it is violating the big-rip theory; i.e. we know that this $a(t)$ should increase with time for our cosmically accelerating universe (specially in phantom era), but FSLLL I do not let it increase, rather, in future this parametrization is making $a(t)$ finite.


Now, we shall study about the nature of the expansion of the universe with FSLL II type of DE model. Again, here we use the EoS of this parametrization. From equation $(9)$ we obtain
\begin{equation}
m_0a^{-3}=M=\rho^{\frac{1}{1+\omega_0+\omega_1\frac{z^2}{1+z^{2}}}}\Rightarrow \rho=(m_0a^{-3})^{(1+\omega_0+\omega_{1}\frac{z(1+z)}{1+z^{2}})}
\end{equation}.
Again using equations $(5)$ and $(7)$
\begin{eqnarray}
\dot{a}^2=2\left\lbrace \frac{4\pi}{3}(m_0a^{-3})^{\left\lbrace 1+\omega_0+\omega_{1}\frac{z^2}{1+z^{2}}\right\rbrace }+\frac{\Lambda}{6}\right\rbrace a^2-k_0 \nonumber \\\Rightarrow \left(\frac{da}{dt}\right)=\left[2\left\lbrace \frac{4\pi}{3}(m_0a^{-3})^{\left\lbrace 1+\omega_0+\omega_{1}\frac{z^2}{1+z^{2}}\right\rbrace }+\frac{\Lambda}{6}\right\rbrace a^2-k_0 \right]^{\frac{1}{2}} 
\end{eqnarray}.
After integrating $t$ with respect to $a(t)$ we get
\begin{eqnarray}
t-t_0=\int_{a(0)}^{a} \left[2\left\lbrace \frac{4\pi}{3}(m_0a^{-3})^{\left\lbrace 1+\omega_0+\omega_{1}\frac{z^2}{1+z^{2}}\right\rbrace }+\frac{\Lambda}{6}\right\rbrace a^2-k_0 \right]^{-\frac{1}{2}}da \label{eq3}
\end{eqnarray}
To obtain the analytic solution of $(22)$,we put $\omega_0=-1$, $\omega_1=0$, $m_0=1$ and get:\\For $k_0=0$
\begin{equation}
t(a)= \sqrt{\frac{3}{8\pi-1}}ln a
\end{equation}
\\For $k_0=1$ 
\begin{equation}
t(a)= \sqrt{\frac{3}{8\pi-1}}ln \left( -2a\sqrt{8\pi-1}+2\sqrt{a^2(8\pi-1)-3}\right) 
\end{equation}
\\For $k_0=-1$ 
\begin{equation}
t(a)= \sqrt{\frac{3}{8\pi-1}} \left(sinh^{-1}\left[\frac{a\sqrt{8\pi-1}}{\sqrt{3}} \right] \right)
\end{equation}

Solving the equation (\ref{eq3}) numerically we plot graphs of $t$ vs $a(t)$ for $k_0=1, 0, -1$. For $k_{0}=1$, we get the graph $5(a)$. Here we compare the case for quintenssence and phantom era with Barboza Alcaniz parametrization for different values of $k$ and $\omega$. 
\begin{figure}[ht!]
\begin{center}
$~~~~Fig.5(a)~~~~~~~~~~~~~~~~~~~~~~~~~~~~~~~~~~~~~Fig.5(b)~~~~~~~~~~~~~~~~~~~~~~~~~~~~~~~5(c)$
\includegraphics[height=5cm, width=5.6cm]{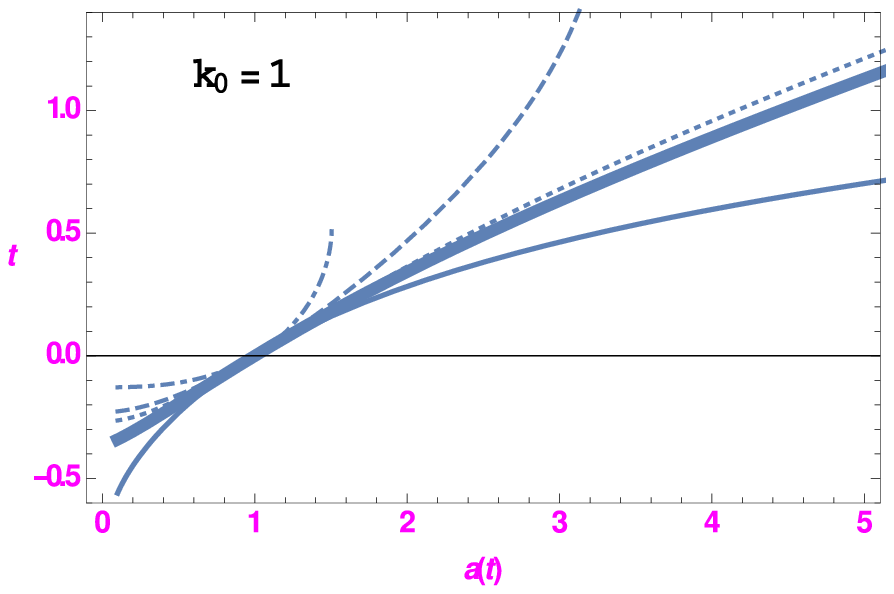}~~~~~\includegraphics[height=5cm, width=5.6cm]{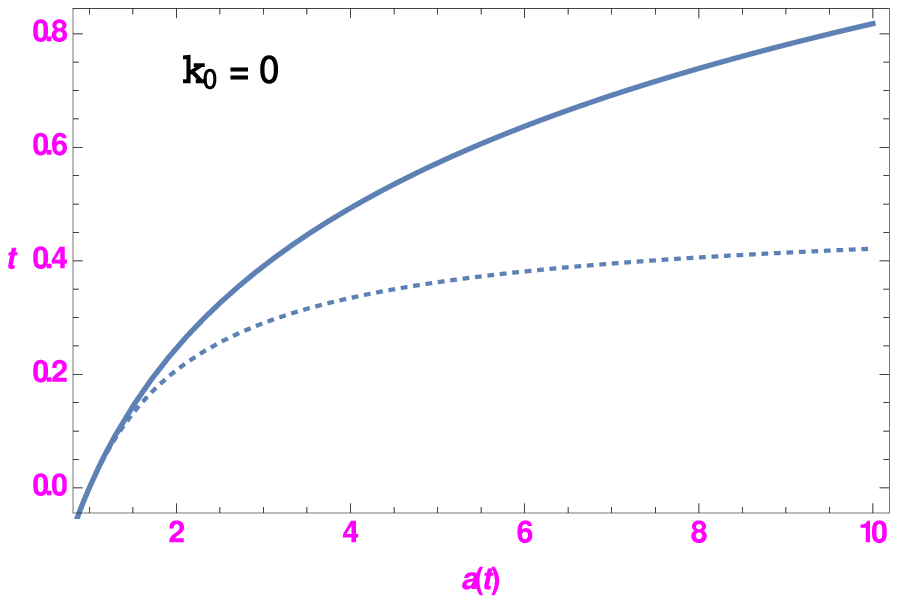}~~~\includegraphics[height=5cm, width=5.6cm]{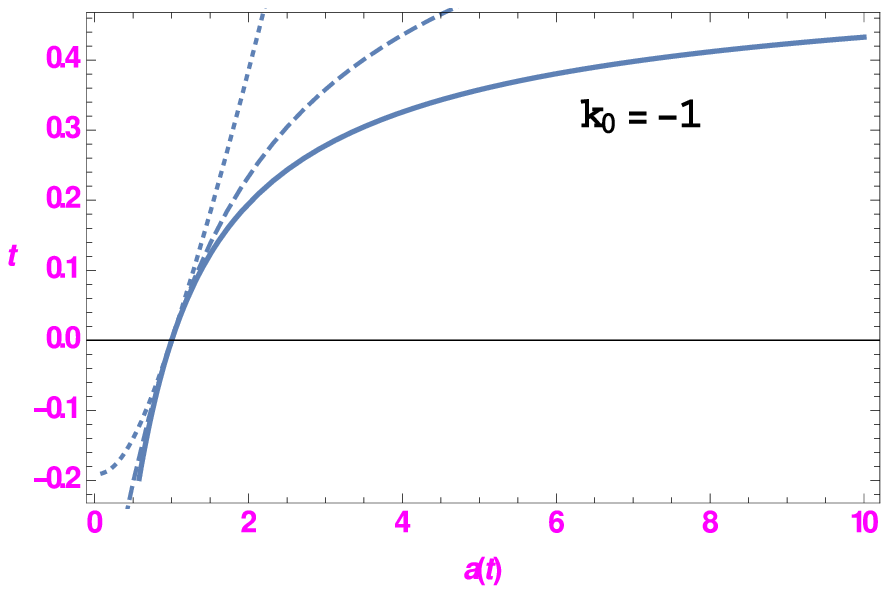}
\end{center}
Figure $5(a)$ to $5(c)$ are $t$ vs. $a(t)$ plots of FSLL II parametrization for quintessence era for $k_0=1$, $k_0=0$ and $k_0=-1$ respectively. For Fig.$5(a)$: solid line stands for $\omega_0=-0.9$, $\omega_1=0.7$; dotted line represents $\omega_0=-0.60423$, $\omega_1=1.40845$; the dot-dashed line stands for $\omega_0=1.006$ and $\omega_1=-0.41493775933$; dashed line represents $\omega_0=-0.2$, $\omega_1=0.7$ and the thicker one represents $\omega_0=-0.6$, $\omega_1=0.7$. For Fig.$5(b)$: solid line stands for $\omega_0=-1$, $\omega_1=0.1$ and the dotted line represents $\omega_0=-1.50423$, $\omega_1=1.40845$. For Fig.$5(c)$: the solid line stands for $\omega_0=-1.4$, $\omega_1=0.1$; dotted line represents $\omega_0=-0.30423$, $\omega_1=1.40845$; the dashed one stands for $\omega_{0}=-1.20423,\omega_{1}=1.40845$ and the dot-dashed one represents $\omega_{0}=0.5170224481,\omega_{1}=-0.3149677893$.
\end{figure}

The graphs of FSLL II parametrization change rapidly for closed universe ($k_0=1$). Here we get totally different graphs for same $\omega_1$ ($0.7$) making a little change in $\omega_0$ (the thick, dashed and more thicker curves). The thick line is purely divergent, the thicker line is slightly divergent but the dashed line converges with increasing $t$. The dotted line ($\omega_0=-0.60423$, $\omega_1=1.40845$) gives no proper conclusion and the dot-dashed line ($\omega_0=1.006$ and $\omega_1=-0.41493775933$) becomes asymptotic to a finite value for increasing $t$ is absent.

For flat universe ($k_0=0$), the graphs become divergent with the increment of $a(t)$. The range of another graph suddenly rises so high ($10^{30}$) that no such comparison can be made and we have omitted it from the figure.

In case of open universe we see that the dotted line ($\omega_0=-0.30423$, $\omega_1=1.40845$) is purely convergent, the dashed line ($\omega_{0}=-1.20423,\omega_{1}=1.40845$) gives no conclusion and the solid line ($\omega_0=-1.4$, $\omega_1=0.1$) leads to divergence.

\begin{figure}[ht!]
\begin{center}
$~~~~~~~~~~Fig.6(a)~~~~~~~~~~~~~~~~~~~~~~~~~~~~~~~~~~~~~~~~Fig.6(b)~~~~~~~~~~~~~~~~~~~~~~~~~~~~~~~~~~~~~~~~~Fig.6(c)$
\includegraphics[height=5cm, width=5.6cm]{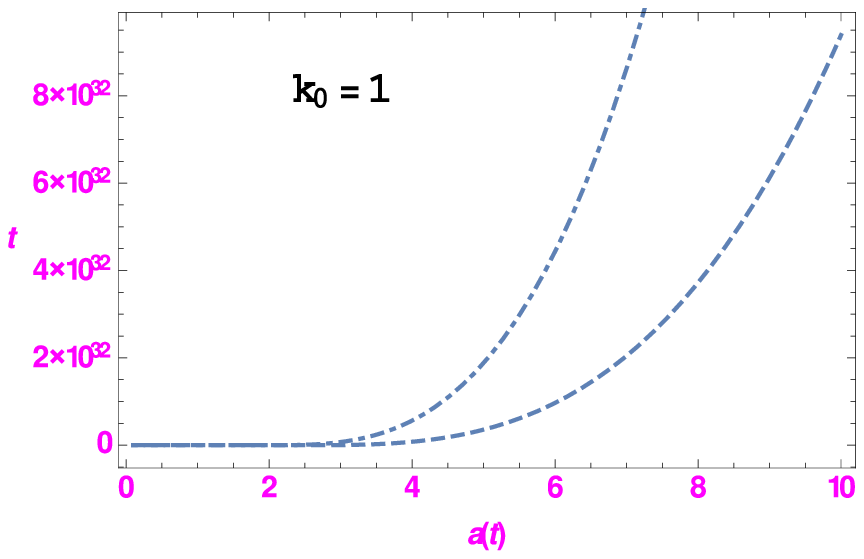}~~~~~~~\includegraphics[height=5cm, width=5.6cm]{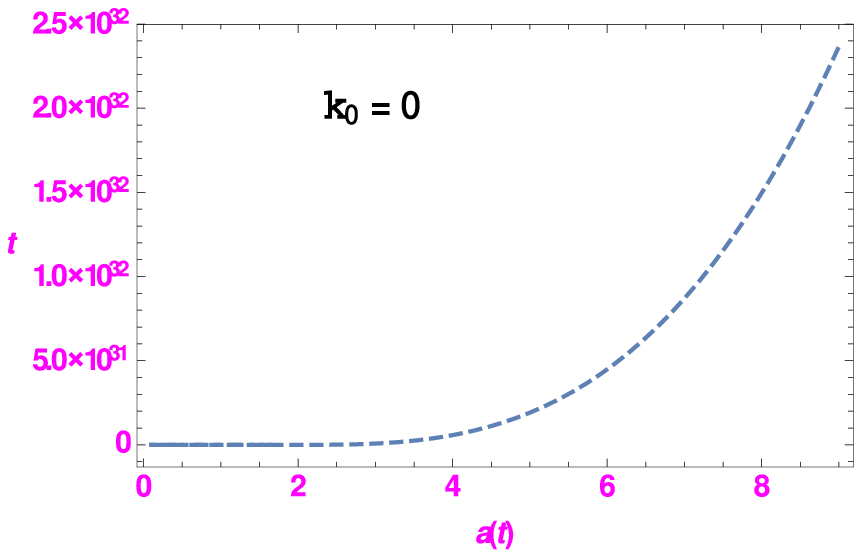}~~~~~~~~\includegraphics[height=5cm, width=5.6cm]{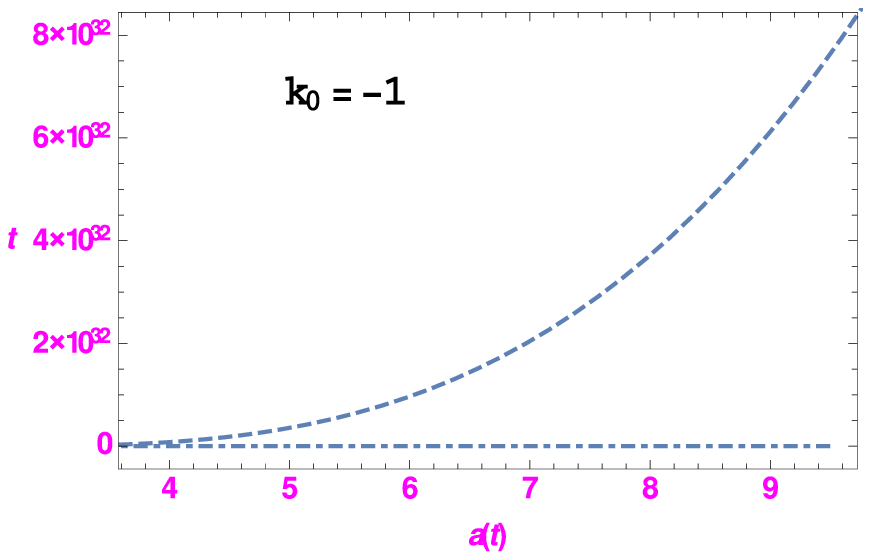}
\end{center}
Figure $6(a)$ to $6(c)$ are $t$ vs. $a(t)$ plots of FSLL II parametrization for phantom era for $k_0=1$, $k_0=0$ and $k_0=-1$ respectively. For Fig.$6(a)$: dashed line represents $\omega_0=0.5$, $\omega_1=-0.799$ and the dot-dashed one stands for $\omega_0=1.2$ and $\omega_1=0.9$. For Fig.$6(b)$: dashed line represents $\omega_0=0.8$, $\omega_1=-1.895$ and the dot-dashed one stands for $\omega_{0}=1.1,\omega_{1}=0.6$. For Fig.$6(c)$: dashed line represents $\omega_0=0.5$, $\omega_1=-0.799$ and the dot-dashed one represents $\omega_0=1.1$, $\omega_1=0.9$.
\end{figure}

In phantom era, for closed universe ($k_0=1$) the dashed ($\omega_0=0.5$, $\omega_1=-0.799$) and dot-dashed ($\omega_0=1.2$ and $\omega_1=0.9$) lines become asymptotic to a finite value  for highly increasing $t$ (in range of $10^{30}$. 

In fig. $6(b)$ we notice that the dot-dashed line ($\omega_{0}=1.1,\omega_{1}=0.6$) is quiet similar to the dashed one ($\omega_0=0.8$, $\omega_1=-1.895$) and coincides with it after a little increment of $a(t)$ although the values of corresponding $\omega_1$ of the curves are enough different.

In case of open universe, we note that both the dashed ($\omega_0=0.5$, $\omega_1=-0.799$) and dot-dashed ($\omega_0=1.1$, $\omega_1=0.9$) lines are divergent with high $t$. The dot-dashed one becomes parallel with the axis of $a(t)$ for increasing $t$.

We observe that the graphs for $k_0= 1, 0, -1$, $a(t)$ are not diverging to an infinite value for increasing $t$ in quintessence era but in phantom era they are divergent. The rate of divergence increases for flat universe than the case of closed universe and finally for open universe it is totally divergent.


Lastly, we use a DE model of type polynomial parametrization which is a bit different from other parametrizations viz. BA, FSLL I and FSLL II discussed earlier in this letter. Now using the EoS of this redshift parametrization, in equation $(9)$ we have introduced $a(t)$ vs $t$ graphs.
\begin{equation}
m_0a^{-3}=M=\rho^{\frac{1}{(1+\omega_0)(\frac{1+2z}{1+z})}}\Rightarrow \rho=(m_0a^{-3})^{(1+\omega_0)(\frac{1+2z}{1+z})})
\end{equation}.
Note that this type of parametrization is somewhat different from previous cases as it depends only upon the values of $\omega_0$ ($\omega_1$ vanishes). Again using equations $(5)$ and $(7)$ we obtain
\begin{eqnarray}
\dot{a}^2=2\left\lbrace \frac{4\pi}{3}(m_0a^{-3})^{\left\lbrace (1+\omega_0)(\frac{1+2z}{1+z}) \right\rbrace }+\frac{\Lambda}{6}\right\rbrace a^2-k_0 \nonumber \\\Rightarrow \left(\frac{da}{dt}\right)=\left[2\left\lbrace \frac{4\pi}{3}(m_0a^{-3})^{\left\lbrace (1+\omega_0)(\frac{1+2z}{1+z})\right\rbrace }+\frac{\Lambda}{6}\right\rbrace a^2-k_0 \right]^{\frac{1}{2}} 
\end{eqnarray}.
Here, integrating $t$ with respect to $a(t)$ we get
\begin{eqnarray}
t-t_0=\int_{a(0)}^{a} \left[2\left\lbrace \frac{4\pi}{3}(m_0a^{-3})^{\left\lbrace (1+\omega_0)(\frac{1+2z}{1+z})\right\rbrace }+\frac{\Lambda}{6}\right\rbrace a^2-k_0 \right]^{-\frac{1}{2}}da \label{eq4}
\end{eqnarray}.
To obtain the analytic solution of $(28)$,we put $\omega_0=-1$, $\omega_1=0$, $m_0=1$ and get:\\For $k_0=0$
\begin{equation}
t(a)= \sqrt{\frac{3}{8\pi-1}}ln a
\end{equation}
\\For $k_0=1$ 
\begin{equation}
t(a)= \sqrt{\frac{3}{8\pi-1}}ln \left( -2a\sqrt{8\pi-1}+2\sqrt{a^2(8\pi-1)-3}\right) 
\end{equation}
\\For $k_0=-1$ 
\begin{equation}
t(a)= \sqrt{\frac{3}{8\pi-1}} \left(sinh^{-1}\left[\frac{a\sqrt{8\pi-1}}{\sqrt{3}} \right] \right)
\end{equation}

Solving the equation (\ref{eq4}) numerically we plot graphs of $t$ vs $a(t)$ for $k_0=0,1,-1$. For $k_{0}=1$, we get the graph of $7(a)$. The case for quintenssence and phantom era for different values of $k$ and $\omega$ has been discussed here.
\begin{figure}[ht!]
\begin{center}
$~~~~Fig.7(a)~~~~~~~~~~~~~~~~~~~~~~~~~~~~~~~~~~~~~Fig.7(b)~~~~~~~~~~~~~~~~~~~~~~~~~~~~~~~~~~~~~~7(c)$
\includegraphics[height=5cm, width=5.6cm]{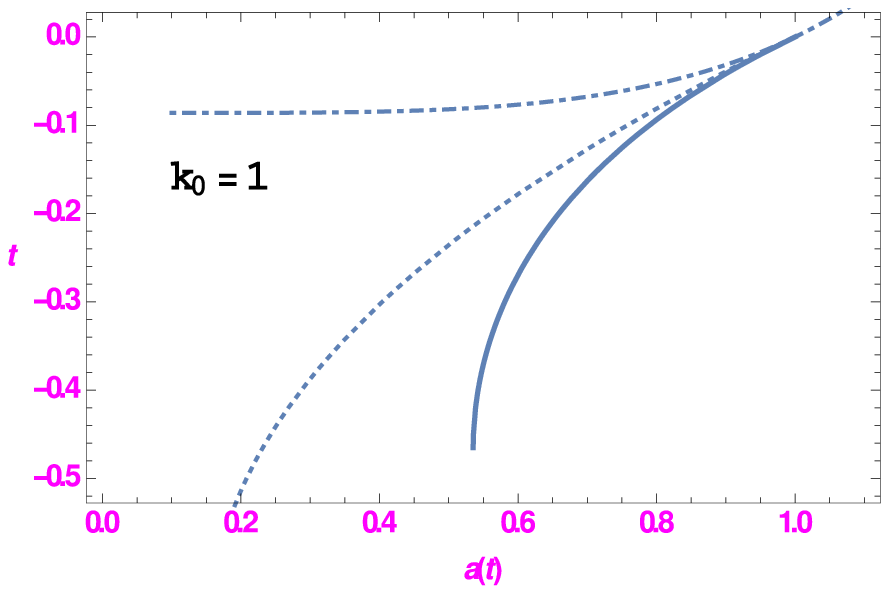}~~~~~\includegraphics[height=5cm, width=5.6cm]{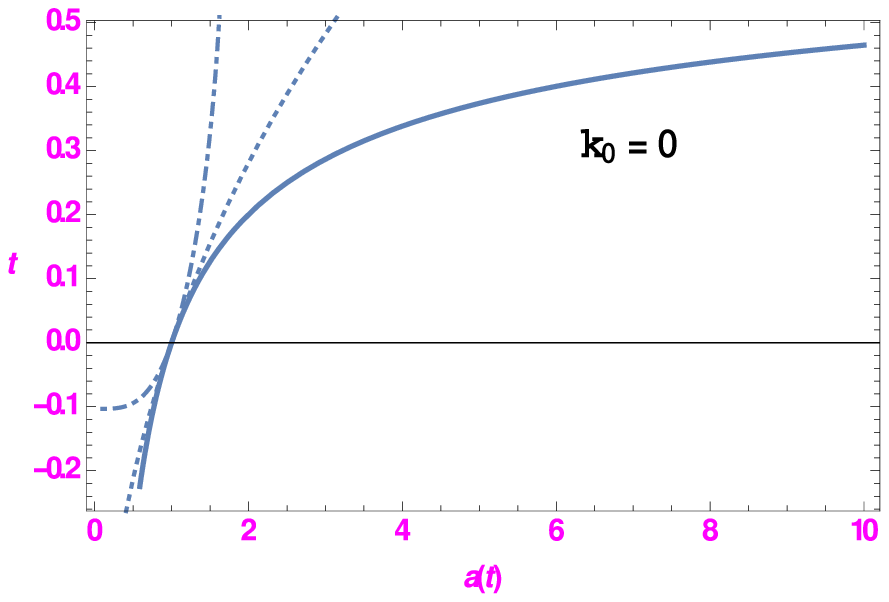}~~~~~\includegraphics[height=5cm, width=5.6cm]{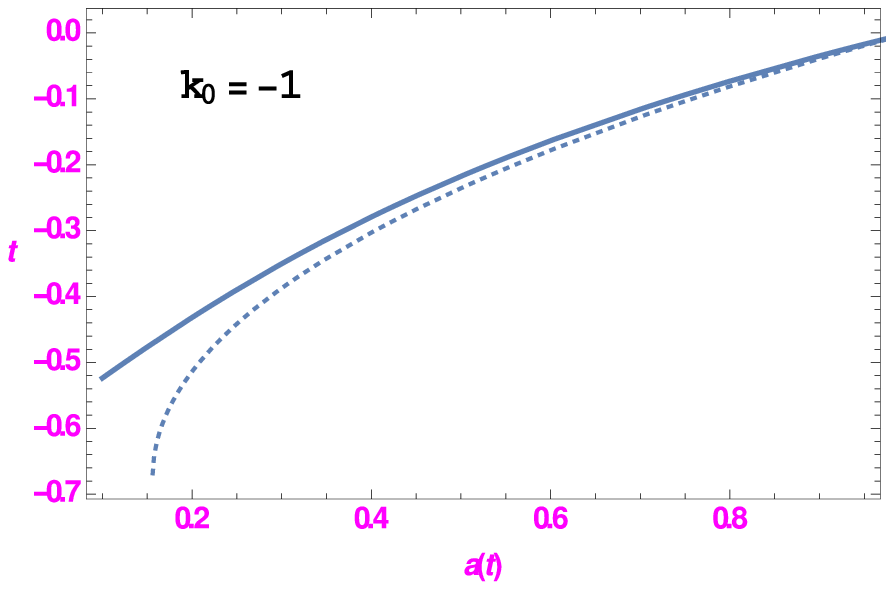}
\end{center}
Figures $7(a)$ to $7(c)$ are $t$ vs. $a(t)$ plots of polynomial parametrization for quintessence era for $k_0=1$, $k_0=0$ and $k_0=-1$ respectively. For Fig.$7(a)$: the solid line stands for $\omega_0=-1.3$; dotted line represents $\omega_0=-0.80423$ and the dot-dashed one stands for $\omega_0=1.006$. For Fig.$7(b)$: solid line stands for $\omega_0=-1.3$; dotted line represents $\omega_0=-0.80423$ and the dot-dashed one stands for $\omega_{0}=0.6170224481$. For Fig.$7(c)$: solid line stands for $\omega_0=-1$ and the dotted line represents $\omega_0=-0.80423$. 
\end{figure}

In closed universe (fig. $7(a)$) the graphs state that for negative time we get an increasing negative $a(t)$; which represents deceleration in past time. Here all the solid ($\omega_0=-1.3$), dotted ($\omega_0=-0.80423$) and dot-dashed ($\omega_0=1.006$) lines coincide just below the axis of $a(t)$ near $t=0.05$.

We see that $a(t)$ blows with the increment of $t$ for flat universe ($k_0=0$). Here, the divergent solid line ($\omega_0=-1.3$) is asymptotic to a finite value of $a(t)$. But the dotted ($\omega_0=-0.80423$) and dot-dashed lines ($\omega_{0}=0.6170224481$) converge with increasing $t$. The solid line does not allow any future cosmological singularity.

In Figure $7(c)$ we observe that the polynomial parametrization gives quiet similar curves for open universe ($k_0=-1$). Here we get solid ($\omega_0=-1$) and dotted ($\omega_0=-0.80423$) lines in negative region. The lines coincide near the axis of $a(t)$ and from their behaviour we state that deceleration has been observed in past.

\begin{figure}[ht!]
\begin{center}
$~~~~~~~~~~~~~~~~~~~~~~~~Fig.8~~~~~~~~~~~~~~~~~~~~~~~~~~~~~~~~~~~~~~$\\
\includegraphics[height=5cm, width=5.6cm]{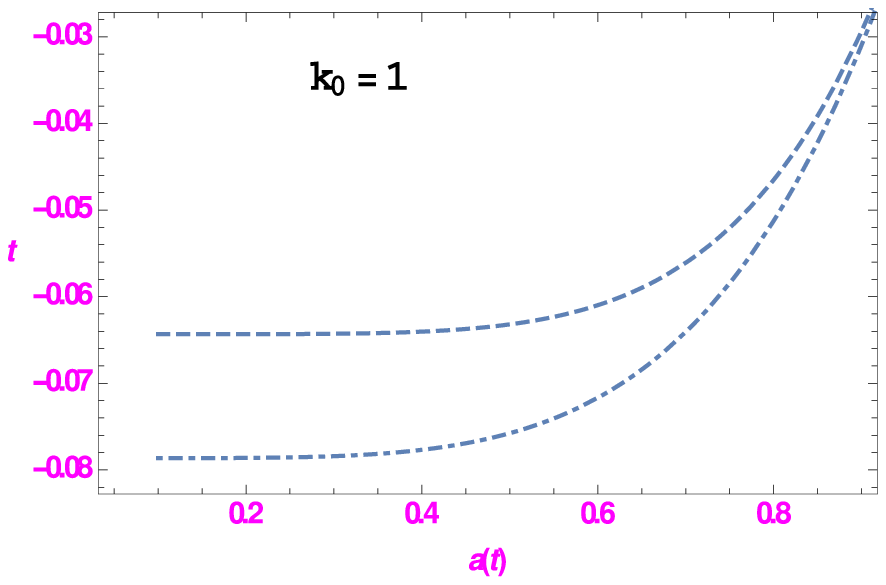}~~~~~~~~~~~~~~~~~~~~
\end{center}
Figure $8$ is a $t$ vs. $a(t)$ plot of polynomial parametrization for phantom era for $k_0=1$ only. Here, the dashed line represents $\omega_0=1.7$ and the dot-dashed one stands for $\omega_0=1.2$ .
\end{figure}
In this figure $8$, both the dotted ($\omega_0=1.7$) and the dashed lines ($\omega_0=1.2$) give increasing graphs in negative region (it seems that for negative time we get a negative but increasing $a(t)$) i.e. in past we may observe a deceleration. 

In polynomial parametrization, we have found a graph only for $k_0=1$ and the rest for $k_0=0$ and $k_0=-1$ are either unphysical, or coincide with the above graph, or their range suddenly rises so high that no such important comparison can be concluded. In this parametrization $\omega_1$ vanishes and so it depends only upon the values of $\omega_0$. Here it should be pointed out that for the values of $\omega$, the values of $a(t)$ converge to a finite value for increasing $t$. Unlike other parametrizations, here the graphs often coincide near the axis of $a(t)$ for small $t$.

Finally we will discuss in brief regarding the results found in this letter. In this letter, study of the evolution of scale factor $a(t)$ and time $t$ has been done. Equations of state of various redshift parametrization have been taken and some comparative study of some important redshift parametrizations has been studied. Firstly, we have considered the FLRW metric and using mainstream families of redshift parametrization, stated EoS of different parametrizations. Here we have used some important well-known equations like Einstein's field equation, energy conservation equation, Raychoudhuri equation etc. and did some preliminary calculations. Here a mass function $M(\rho)$ has been introduced and we put the EoS of BA, FSLL I, FSLL II, polynomial parametrizations in the mass function and solved differential equations of time ($t$) vs. scale factor ($a(t)$) numerically and plotted graphs with the values of $\omega_0$, $\omega_1$ for closed, flat and open universes ($k_0=1,0,-1$). The simple looking graphs have enormous effect upon the hypothesis of the expansion of the universe. Now if we think about the increment of $a(t)$ with respect to $t$ then from the graphs we conclude that for closed universe ($k_0=1$) $a(t)$ becomes parallel with the axis of $t$ after a certain range of values of $t$ and $a(t)$. For the case of flat universe ($k_0=0$) the incidents are quite similar. In case of open universe ($k_0=-1$); this $a(t)$ becomes parallel after a little increment of $t$  i.e. we get a finite $a(t)$. Here we get a new result from BA parametrization. Where $a(t)$ should be very large and somehow the universe should accelerate abruptly and will burst out (Big-Rip), BA states that in phantom era $a(t)$ becomes finite for some particular $t$.

So we can conclude in brief that in a hypothetical cosmological model Big Rip concerns the ultimate fate of the universe, in which the matter of the universe and even space-time itself is progressively torn apart by the expansion of the universe. the universe dominated by phantom energy is an accelerating universe, expanding at an ever-increasing ratio. When the size of the observable universe became smaller than any particular structure, no interaction by any of the fundamental forces (gravitational, electromagnetic, strong and weak) can occur between the most remote part s of the structure. When these interactions become impossible, the structure is ripped apart. 

In case of FSLL I and polynomial parametrizations; the same pattern is followed for all the cases of closed, flat and open universe ($k_0=1, 0, -1$). These parametrizations are not allowing infinite $a(t)$ even when the parameters are signifying phantom era. However FSLL II parametrization somehow supports the big-rip hypothesis. \\
\textbf{Acknowledgement}

This research is supported by the project grant of Government of West Bengal, Department
of Higher Education, Science and Technology and Biotechnology (File no:- ST /P/S\&T /16G − 19/2017). RB thanks IUCAA, Pune for Visiting Associateship. RB dedicates this article to his PhD supervisor Prof. Subenoy Chakraborty, Department of Mathematics, Jadavpur University, Kolkata-32, India to tribute him on his $60^{th}$ birth year.

\end{document}